\documentclass[openacc]{rstransa}%%%%where rstrans is the template name

\begin{document}

%%%% Article title to be placed here
\title{Quantum information approach to\\ high energy interactions}

\author{%%%% Author details
Dmitri E. Kharzeev$^{1,2}$}

%%%%%%%%% Insert author address here
\address{$^{1}$Center for Nuclear Theory,\\ Department of Physics and Astronomy,\\
Stony Brook University, New York 11794-3800
$^{2}$Physics Department, Brookhaven National Laboratory, Upton, New York 11973-5000}

%%%% Subject entries to be placed here %%%%
\subject{High energy physics, Quantum information, Strong interactions}

%%%% Keyword entries to be placed here %%%%

%%%% Insert corresponding author and its email address}
\corres{Dmitri Kharzeev\\
\email{dmitri.kharzeev@stonybrook.edu}}

%%%% Abstract text to be placed here %%%%%%%%%%%%
\begin{abstract}
High energy hadron interactions are commonly described by using a probabilistic parton model that ignores quantum entanglement present in the light-cone wave functions. Here we argue that since a high energy interaction samples an instant snapshot of the hadron wave function, the phases of different Fock state wave functions cannot be measured -- therefore the light-cone density matrix has to be traced over these unobservable phases. Performing this trace with the corresponding $U(1)$ Haar integration measure leads to ``Haar scrambling" of the density matrix, and to the emergence of entanglement entropy. This entanglement entropy is determined by the Fock state probability distribution, and is thus directly related to the parton structure functions. As proposed earlier, at large rapidity $\eta$ the hadron state becomes maximally entangled, and the entanglement entropy is $S_E \sim \eta$ according to QCD evolution equations. When the phases of Fock state components are controlled, for example in spin asymmetry measurements, the Haar average cannot be performed, and the probabilistic parton description breaks down.
\end{abstract}
%%%%%%%%%%%%%%%%%%%%%%%%%%%

%%%%%%%%%% Insert the texts which can accomdate on firstpage in the tag "fmtext" %%%%%
\newpage
\begin{fmtext}

\end{fmtext}

%%%%%%%%%%%%%%% End of first page %%%%%%%%%%%%%%%%%%%%%

\maketitle
\section{Introduction}
Quantum information emerges as a new universal language for describing the behavior of a wide variety of quantum systems; for recent reviews, see \cite{Preskill:2021apy,Klco:2021lap}. Here, we will attempt to use this language for the description of high energy interactions, with the hope of getting new insights on the long-standing problem of describing hadrons in terms of their fundamental constituents.

 Hadrons -- the bound states of quarks and gluons -- store information about the dynamics of strong interactions. An access to this information is provided by high energy collisions in which detectors of produced final states serve as information readout devices. This readout process is necessarily accompanied by the degradation of information content, and thus the creation of Shannon entropy
 \begin{equation}\label{shan}
 H = - \sum_n p_n \ln p_n ,
 \end{equation}
 where $p_n$ is the probability of a particular measurement outcome taken from the space of all possible outcomes, with $\sum_n p_n =1$\footnote{Using natural logarithm instead of $\log_2$ in (\ref{shan}) amounts to counting entropy in ``nats" instead of bits.}. From the physics viewpoint, the entropy (\ref{shan}) represents the Gibbs entropy of the produced final state.
The degradation of information content is not (only) due to detector imperfections, but is a fundamental consequence of the space-time picture of high energy interactions that does not allow to recover all information about the hadron wave function in the measurement \cite{Kharzeev:2017qzs}.  

 For example, the process of Deep Inelastic Scattering (DIS) samples a part of the wave function of the proton in the vicinity of the light cone. The proton represents a pure quantum state described by the density matrix ${\hat \rho}$ with zero von Neumann entropy. However, the state measured by DIS is characterized by the density matrix ${\hat \rho_{\rm m}}$ that is traced over the degrees of freedom $nm$ not accessible for the observation:
 \begin{equation}\label{mix}
 {\hat \rho_{\rm m}} = {\rm Tr}_{\rm nm} {\hat \rho} .
 \end{equation}
 This density matrix describes a mixed state with non-zero von Neumann entropy 
 \begin{equation}
 S_E = - {\rm Tr}\ {\hat \rho_{\rm m}} \ln {\hat \rho_{\rm m}} .
 \end{equation}
 
 The Schmidt decomposition theorem guarantees the existence of a basis $| i \rangle$ in which the density matrix (\ref{mix}) can be written down as a {\it single} sum:
  \begin{equation}\label{mix_s}
 {\hat \rho_{\rm m}} = \sum_n p_n | n \rangle \langle n |,
 \end{equation}
where $p_n$ is the probability to find the state $| n \rangle$ in the incoherent superposition (\ref{mix_s}). The corresponding von Neumann entropy is
\begin{equation}\label{vonN}
 S_E = - \sum_n p_n \ln p_n .
 \end{equation}
It seems natural to identify the Schmidt basis states with the states detected in an experiment -- if we do this, then the Gibbs (Shannon) entropy of the final state (\ref{shan}) becomes equal to the entanglement entropy of the initial state (\ref{vonN}), $S_E = H$.  Indeed, a recent computation \cite{Florio:2021xvj} provided an explicit example of this equality by establishing the equality of the entanglement and Gibbs entropies in pair production by electric pulses. 

In view of these considerations, it becomes important to identify the physical meaning of the Schmidt basis $| n \rangle$ in (\ref{mix_s}) in hadronic interactions. It has been argued in \cite{Kharzeev:2017qzs} that this basis in high energy interaction is provided by Fock states with different numbers of partons $n$. The Gibbs entropy of the produced system of hadrons is evaluated from the multiplicity distribution of the produced hadrons, with $p_n$ representing the probability to detect $n$ hadrons.
The equality of the entanglement and Gibbs entropies would thus imply a stronger, event-by-event, version of the ``local parton-hadron duality" \cite{Dokshitzer:1987nm} between the number of partons and  the multiplicity of the produced hadrons. 

In the present paper, we explore the quantum information interpretation of parton model further. Because the field theoretic basis of the parton model is provided by the light-cone quantization, we will seek to identify the degrees of 
freedom that are not observed on the light cone, find the density of matrix of the observed state by tracing over the unobserved degrees of freedom, and evaluate the corresponding entanglement entropy. We will show that this entropy originates from the entanglement between the observed (parton number) and unobserved (phase) degrees of freedom on the light cone.

\section{Light-cone wave functions}\label{sec:lc}

The idea of light-cone quantization dates back to Dirac \cite{Dirac:1949cp}, who observed that quantum dynamics 
on the three-dimensional surface $t - z = 0$ in space-time formed by a plane wave front of light (``front form dynamics") may appear more simple than the traditional ``instant form dynamics". This is because in front form dynamics, seven (of ten)  Poincar\'e group generators, including a particular boost generator, leave the light front invariant and are independent of the Hamiltonian, whereas in the instant form only six Poincar\'e generators (translations and rotations) leave the instant surface $t=0$ invariant, and the remaining four (translations in time and boosts) involve the Hamiltonian. This feature of the light-cone dynamics enables a frame-independent description of the wave functions. 

The light-cone quantization emerged as a useful tool to describe hadrons in terms of their quark and gluon constituents. 
The hadron wave function on the light cone (for a comprehensive review and references, see \cite{Brodsky:1997de}) is given by 
\begin{equation}\label{lc}
|\Psi \rangle = \sum_n^\infty \int d\Gamma_n\ \Psi_n(x_i, {\vec k}_{\perp i}) \prod_i^n a^\dagger_i (x_i, {\vec k}_{\perp i}) |0\rangle, 
\end{equation}
where $d \Gamma_n$ is the phase space differential
\begin{equation}\label{ps}
d\Gamma_n =  \delta(1-\sum_i^n x_i) \delta^{(2)}(\sum_i^n {\vec k}_{\perp i}) dx_1 ... dx_n d {\vec k}_{\perp 1} ... d {\vec k}_{\perp n} ,
\end{equation}
$a^\dagger_i (x_i, {\vec k}_{\perp i})$ are the parton creation operators, and $\Psi_n(x_i, {\vec k}_{\perp i})$ are the light-cone wave functions of Fock states with $n$ partons. 

Note that $\Psi_n$ describes the entanglement between partons. Yet, virtually all applications of light-cone wave functions deal with probability distributions $|\Psi_n(x_i, {\vec k}_{\perp i})|^2$ and ignore quantum interference between the states with different numbers of partons. For example, the gluon structure function at Bjorken $x$ and scale $Q^2$ is given by 
\begin{equation}\label{glu}
G(x, Q^2) = \sum_n^\infty \int d\Gamma_n\ \sum_{i \in g} \delta(x-x_i)\ |\Psi_n(x_i, {\vec k}_{\perp i})|^2 , 
\end{equation}
where the integration over the transverse momenta of the gluons is performed up to $Q$, the resolution scale of the probe. 

This probabilistic formula expresses the essence of the parton model -- partons are treated as free particles characterized by probability distributions in momentum and/or coordinate space. At first glance this looks natural due to the asymptotic freedom of QCD -- at short distances, quarks and gluons begin to behave as free particles. There is a big problem however -- 
an ensemble of free particles is characterized by non-zero entropy resulting from the different possible positions in phase space. At the same time the hadron wave function (\ref{lc}) is a pure state with zero entropy. What is the origin of the parton entropy?

\section{Information scrambling in high-energy interactions}\label{sec:scramble}

In this section we will introduce the idea of high-energy information scrambling by using first a familiar ``instant" form of Hamiltonian dynamics, in Dirac's classification \cite{Dirac:1949cp}. Let us begin by considering the hadron in its rest frame probed by a pulse of light. 
The Schroedinger equation describing the hadron is 
\begin{equation}
i \frac{\partial}{\partial t} | \Psi \rangle = {\hat H} | \Psi \rangle ;
\end{equation}
the hadron is an eigenstate of the QCD Hamiltonian in the rest frame:
\begin{equation}
{\hat H} | \Psi \rangle = M | \Psi \rangle,
\end{equation}
with eigenvalue corresponding to the hadron mass $M$.

Let us now define Fock states with a fixed number of partons $n$:
\begin{equation}\label{fock}
| n \rangle =  \frac{1}{\sqrt{n!}} \prod_i^n a^\dagger_i  |0\rangle ,
\end{equation}
that we will assume to be eigenstates of a Fock oscillator Hamiltonian ${\hat H}_F$ with energy eigenvalues $E_n = \omega (n + 1/2)$:
\begin{equation}
{\hat H}_F | n \rangle = E_n |n \rangle .
\end{equation}
Since Fock states represent a complete basis, we can
represent the ground hadron state $| \Psi_0 \rangle$ as their superposition:
\begin{equation}\label{fock_rest}
| \Psi \rangle = \sum_n \alpha_{n}  |n\rangle ,
\end{equation}
where $\alpha_{n} = \langle n | \Psi \rangle$ is the rest-frame analog of the light-cone wave function (\ref{lc}).

The density matrix of the hadron is 
\begin{equation}\label{rho-h}
{\hat \rho} = | \Psi \rangle  \langle \Psi \rangle = \sum_{n, n'} \alpha_{n}\ \alpha_{n'}^*\ | n \rangle  \langle n' | ,
\end{equation}
which obviously represents a pure state. Each Fock state evolves in time according to $|n(t) \rangle = \exp(-i E_n t) |n(t) \rangle$ with $E_n = \omega (n + 1/2)$, so the time evolution of this density matrix is described by
\begin{equation}\label{rho-t}
{\hat \rho}(t)  = \sum_{n,n'} e^{i (n' - n) \omega t}\ {\hat \rho}(t=0). 
\end{equation}

Now, suppose that we perform a measurement of the Fock state decomposition in a hadron by a light front. The light front will pass through the hadron in a time $t_{light} = R$, where $R$ is the hadron radius. However to determine the phase of the density matrix (\ref{rho-t})  would take the measurement time $t_{meas} \gg 1/\omega$. The energy cost $\omega$ of adding a parton is on the order of QCD scale $\Lambda_{\rm QCD}$, and the hadron size is $R \simeq \Lambda_{\rm QCD}$ -- therefore, the measurement of the phase of the density matrix in our experiment is impossible. In other words, the light front captures an instant snapshot of the hadron wave function and cannot detect its time evolution.

This means that the density matrix of the pure hadron state (\ref{rho-h}) 
has to be traced over the unobserved phase $\varphi \equiv \omega t$.  The Haar integration measure of the $U(1)$ group is simply $d \varphi / 2\pi$, therefore  we can perform the trace of the density matrix (\ref{rho-h}) over the phase as follows: 
\begin{equation}\label{density_part}
{\hat \rho}_{parton} = {\rm Tr}_\varphi {\hat \rho} = \sum_{n,n'}\ \int_0^{2\pi} \frac{d \varphi}{2 \pi} e^{i(n'-n) \varphi}\  \alpha_{n}\ \alpha^*_{n'} | n \rangle \langle n' | = \sum_n |\alpha_{n}|^2 \ | n \rangle \langle n | .
\end{equation}

Note that while the hadron density matrix (\ref{rho-h}) describes a pure state, the ``Haar scrambled" (the term introduced in \cite{Sekino:2008he})  parton density matrix (\ref{density_part}) is an incoherent, probabilistic, superposition of parton Fock states. The parton state probabilities $p_n \equiv |\alpha_{n}|^2$ are real and non-negative, with $\sum_n p_n =1$. Unless only one Fock state contributes, the parton density matrix
\begin{equation}\label{rho-p}
{\hat \rho}_{parton} = \sum_n p_n \ | n \rangle \langle n |
\end{equation}
is mixed, with purity 
\begin{equation}\label{pur}
\gamma_{parton} = {\rm Tr} (\rho_{parton}^2) = \sum_n p_n^2 < 1.
\end{equation}
The corresponding entanglement entropy is 
\begin{equation}\label{EE}
S_E = - \sum_n p_n \ln p_n 
\end{equation}
and describes entanglement between the number of partons and the phase of the corresponding Fock component of the wave function. Since at high energies, due to the short time of the interaction, the phases become unobservable, the density matrix has to be averaged over the phase with a corresponding Haar measure -- this leads to quantum information scrambling and the emergence of the entanglement entropy (\ref{EE}).

The expectation values of operators ${\hat{\cal{O}}}$ at high energies have to be computed according to 
\begin{equation}\label{ave}
\langle {\hat{\cal{O}}} \rangle = {\rm Tr} ({\hat{\cal{O}}} {\hat \rho}_{parton}) = \sum_n p_n \langle n | {\hat{\cal{O}}} | n \rangle;
\end{equation} 
this recipe corresponds to the computation within the parton model. In particular, evaluating the expectation value of the parton number operator ${\hat{\cal{O}}} = {\hat n} = a^\dagger a$, we get the average number of partons in a hadron (at a particular value of Bjorken $x$ and scale $Q^2$ at which the parton probabilities $p_n$ are evaluated), i.e. the parton structure function:
\begin{equation}
\langle n \rangle = {\rm Tr} ({\hat n}\ {\hat \rho}_{parton}) = \sum_n p_n\ n \ .
\end{equation} 

Note that the dominance of a single Fock state (that would be required for the parton density matrix (\ref{rho-p}) to be pure) is very unlikely in QCD, even at small resolution scale. Indeed, due to spontaneously broken chiral symmetry, the spectrum of QCD contains light Goldstone bosons (pions) that couple to the nucleons. As a result, even at small momentum transfer, the wave function of a physical nucleon contains additional pions, $\Delta$ isobars, etc that manifest themselves in hard exclusive reactions, see e.g. \cite{Bass:1996iq}.

\section{Information scrambling on the light cone}

As discussed in section \ref{sec:lc}, the parton model is defined through the wave functions on the light cone, 
so let us discuss the information scrambling in this reference frame. The underlying physics is of course the same as in section \ref{sec:scramble}, but instead of time $t$ and longitudinal coordinate $z$ we now introduce the light-cone time $x^+$ and light-cone space $x^-$:
\begin{equation}
x^+ = \frac{1}{\sqrt{2}}(t + z), \ \ \ x^- = \frac{1}{\sqrt{2}}(t - z) .
\end{equation}
Likewise, the energy $E$ and longitudinal momentum $p_z$ are combined into light-cone momentum $p^+$ and light cone energy $p^-$:
\begin{equation}
p^+ = \frac{1}{\sqrt{2}}(E + p_z), \ \ \ p^- = \frac{1}{\sqrt{2}}(E - p_z) .
\end{equation}
Partons in the hadron are distributed in $p^+$ and the transverse momentum ${\vec p_\perp}$; for example, 
the light-cone wave function of a free gluon in coordinate representation is given by \cite{Brodsky:1997de}: 
\begin{equation}\label{gluon}
A^a_\mu (x) = \sum_{\lambda} \int \frac{dp^+ d^2 p_\perp}{\sqrt{2 p^+ (2 \pi)^3}} \left[a_p\ \epsilon_\mu(p, \lambda) e^{-i p x} + 
a_p^\dagger\ \epsilon_\mu^*(p, \lambda) e^{i p x}\right],
\end{equation}
where $\lambda$ is the gluon helicity, $\epsilon_\mu (p, \lambda)$ is the polarization vector, $a$ is the color index, and creation and annihilation operators satisfy the local commutation relations
\begin{equation}
[a_p, a_p^{\dagger '}] = \delta(p^+ - p^{+'})  \delta^{(2)}(p_\perp - p_\perp ') \delta_\lambda^{\lambda '} \delta_a^{a '}.
\end{equation}
The light-cone of wave functions of quarks and antiquarks are given by analogous expressions, with Fermi anticommutation relation for the creation and annihilation operators.

The Fock states with a fixed number of partons are obtained according to (\ref{fock}), and form a complete set of states with
\begin{equation}
\sum_n \int d \Gamma_n | n \rangle \langle n | = {\hat 1}.
\end{equation} 
The wave function of a hadron on the light cone can be decomposed in the basis of Fock states, in accord with (\ref{lc}):
\begin{equation}
|\Psi \rangle = \sum_n^\infty \int d\Gamma_n\ \Psi_n(x_i, {\vec k}_{\perp i}) |n \rangle .
\end{equation}
The corresponding density matrix describes the pure hadron state, analogously to (\ref{rho-h}):
\begin{equation}\label{rho-lc}
{\hat \rho} = |\Psi \rangle \langle \Psi | = \sum_{n,n'}^\infty \int d\Gamma_n\ d\Gamma_{n'}\ \Psi_{n'}^*(x_{i'}, {\vec k}_{\perp i'}) \Psi_n(x_i, {\vec k}_{\perp i}) |n \rangle \langle n' | .
\end{equation}
In coordinate representation, the light-cone wave functions are proportional to the product of plane waves of individual partons (see (\ref{gluon})); each of these plane waves has a phase
\begin{equation}
x p \equiv x^\mu p_\mu = \frac{1}{2} (x^+ p^- + x^- p^+) - {\vec x}_\perp {\vec p}_\perp .
\end{equation}

Following the derivation in the previous Section, let us assume that the hadron is probed by a light wave front described by $x^-=0$. We therefore know that the positions $z_i$ and times $t_i$ of interaction of light interaction with individual partons 
satisfy $t_i - z_i = x^-_i = 0$, but $z_i$ and $t_i$ individually 
cannot be determined in this measurement -- therefore, the density matrix (\ref{rho-lc}) has to be traced over $x^+_i$, with corresponding Haar measure. This leads to the factor
\begin{equation}\label{constr}
\int \frac{d x^+}{2\pi} e^{i(P_n^- - P_{n'}^-) x^+} = \delta(P_n^- - P_{n'}^-),
\end{equation} 
where $P_n^- = \sum_{i=1}^n p_i^-$. The constraint $P_n^- = P_{n'}^-$ is clearly satisfied for an arbitrary configuration of partons  for $n=n'$, and thus the Haar-scrambled parton density matrix is given by 
\begin{equation}\label{rho-lcp}
{\hat \rho}_{parton} = {\rm Tr}_{x^+} |\Psi \rangle \langle \Psi | = \sum_{n}^\infty \int d\Gamma_n\  | \Psi_n(x_i, {\vec k}_{\perp i})|^2 |n \rangle \langle n | ,
\end{equation}
in analogy with (\ref{density_part}). This parton model density matrix represents a mixed state, in contrast to the density matrix of a pure state  (\ref{rho-lc}).

Therefore the density matrix measurable in high-energy interactions (where the hadron is probed by an object moving with a velocity close to the velocity of light) is diagonal in the number of partons, and describes an information-scrambled, mixed state. 
 The parton probabilities 
 \begin{equation}
 p_n = \int d\Gamma_n\  | \Psi_n(x_i, {\vec k}_{\perp i})|^2
 \end{equation}
can be specified for a particular value of Bjorken $x$ by introducing a factor of $\sum_i \delta(x-x_i)$. For example, the gluon structure function can be evaluated according to (\ref{ave}) as 
\begin{equation}
G(x, Q^2) = {\rm Tr} (a^\dagger a\ {\hat \rho}_{parton}) = \sum_n^\infty \int d\Gamma_n\ \sum_{i \in g} \delta(x-x_i)\ |\Psi_n(x_i, {\vec k}_{\perp i})|^2,
\end{equation}
which coincides with the standard parton model expression (\ref{glu}).

\section{Entanglement entropy at high energies}

At high energies, when multi-parton configurations dominate the hadron wave function, the purity of the mixed parton state (\ref{pur}) is small, and the entanglement entropy (\ref{EE}) is large.  This is due to the increasing complexity of the hadron wave function at high energy -- quantum fluctuations inside the hadron become long-lived due to the Lorentz boost, and can be resolved by a probe. As a result, the number of Fock state components in the hadron wave function increases with energy, and since their phases cannot be determined in a typical high energy interaction, the amount of information lost due to the averaging over these phases grows with energy as well -- this is reflected in the growth of parton entanglement entropy (\ref{EE}).

Let us try and translate this argument into the prediction for the energy dependence of the entanglement entropy (\ref{EE}). The Haar averaging of the hadron density matrix can be expected to produce the entanglement entropy $S_E \sim \ln N$, where $N$ is the number of partons at a given energy $E$. In QCD evolution \cite{Gross:1973id,Politzer:1973fx}, as described by DGLAP \cite{Gribov:1972ri,Altarelli:1977zs,Dokshitzer:1977sg} or BFKL equations \cite{Kuraev:1977fs,Balitsky:1978ic}, $N$ grows as a power of energy\footnote{This is a consequence of resummation of parton emissions each of which happens with a probability $\sim \alpha \ln E$.}, $N \sim E^\alpha$. As a result, we expect that at high energies
\begin{equation}\label{eey}
S_E(E) \sim \ln E \sim y ,
\end{equation}
where $y$ is the hadron rapidity. If we assume in addition that all parton probabilities are equal (i.e. the state is maximally entangled), then since $\sum_n^N p_n =1$, we get $p_n = 1/N$ and 
\begin{equation}\label{eepart}
S_E = - \sum_n p_n \ln p_n = \ln N.
\end{equation}
This relation suggests at high energies there should be a relation between the entanglement entropy and the parton structure functions. In \cite{Kharzeev:2017qzs} this relation has been studied by using the QCD evolution equations for parton multiplicity distribution $p_n$; it has been found that the relations (\ref{eepart}) and (\ref{eey}) indeed hold starting from rapidity $y \simeq 6$. Linear growth of the entanglement entropy with rapidity has also been found in \cite{Armesto:2019mna,Dvali:2021ooc}.

\section{Beyond the parton model}

The justification of the probabilistic parton model proposed here also enables us to realize the limitations of this approach. Even in an inclusive hard process (when all final states are summed over), there is a potential contribution that arises from quantum interference between states with different numbers of partons. Indeed, the constraint $P^-_n = P^-_{n'}$ arising from the averaging of the density matrix over the Haar measure (see (\ref{constr})) is obviously satisfied for $n=n'$, leading to the mixed-state density matrix (\ref{rho-lcp}). However, in QCD (unlike in the oscillator model of Section \ref{sec:scramble}) it can also be satisfied for $n \neq n'$, leading to quantum interference between the amplitudes with different number of partons. This interference is not taken into account in traditional parton model; its numerical significance can be established by using QCD  on the light cone.

While in an inclusive process the phases of parton states are not measured and have to be averaged over, this is not true for spin-dependent measurements where the spin asymmetries are sensitive to the phases of parton amplitudes (for review, see \cite{Aidala:2012mv}). In this case the final and/or initial states introduce a bias in the average over the phase, and the relations (\ref{density_part}) or (\ref{constr})  cannot be used. Such processes can therefore involve interference between amplitudes involving different numbers of partons, which is beyond the traditional probabilistic parton description. It is tempting to assume that this interference may be at the origin of numerous puzzles arising in the description of spin asymmetries within the parton model. 

One specific example is the time-reversal-odd single spin asymmetry in semi-inclusive DIS, and associated quark Sivers function (see \cite{Aidala:2012mv} for a review). This requires generating an imaginary phase of the light-cone wave function, and one way of generating this is due to rescattering of the struck quark with spectator quarks in the nucleon \cite{Brodsky:2002cx}. However at higher order, such a phase could be generated from the interference between the light-cone wave functions with different number of partons. This interference should arise from evolving the light-cone wave function \cite{Brodsky:2010vs} using the interaction kernel beyond the leading order.

\section{Conclusion and outlook}

To summarize, here we have argued that non-zero entropy of partons is the entanglement entropy between the number and phase in the light-cone wave functions. In high energy interactions the phase of the phase of the wave function cannot be determined, so the density matrix of the hadron has to be averaged over the phase with the corresponding Haar integration measure. This yields a density matrix of a mixed state that is diagonal in parton number, which is the density matrix of the parton model. 
 This approach to parton model allows to understand better its limitations, for example in the description of spin-dependent processes.

It would be interesting to extend this study in at least three different directions.
First, it appears important to investigate the emergence of entanglement entropy under Lorentz boost using a specific model for the wave function of a bound state. The first steps in this direction were made in \cite{Feynman:1971wr,Kim:1990pq}, where the Lorentz boost of a covariant oscillator was considered. In particular, it was argued in \cite{Kim:1990pq} that the Lorentz squeeze of the covariant oscillator can result in the emergence of entropy. Extending these studies to light-cone QCD wave functions would allow to quantify the approach advocated here. 
 
Second, the arguments presented above suggest that the emergence of entanglement entropy on the light cone may be described in terms of geometry. As pointed out by Dirac \cite{Dirac:1949cp}, on the light cone the number of 
Poincar\'e group generators that do not involve the Hamiltonian increases to seven (from six in the traditional ``instant form"). It seems natural to conjecture that the Haar scrambling of the density matrix considered above corresponds to averaging over the orientations in this additional degree of freedom, with the corresponding Haar measure. If this conjecture is correct, then the emergence of entanglement entropy on the light cone is a general phenomenon that should exist even when the parton model itself does not apply (for example, at strong coupling). 

Third, it appears very interesting to explore the real-time dynamics of entanglement entropy after the collision (modeled as a quench), and to describe the transformation of  entanglement entropy to the Gibbs entropy measured in experiment. Early results \cite{Kharzeev:2017qzs,Florio:2021xvj,Baker:2017wtt,Berges:2017zws,Tu:2019ouv} indicate the promise of this approach in describing the observed featured of high energy collisions.

\vskip6pt

\enlargethispage{20pt}

%\ethics{Insert ethics statement here if applicable.}

%\dataccess{Insert details of how to access any supporting data here.}

%\aucontribute{For manuscripts with two or more authors, insert details of the authors’ contributions here. This should take the form: 'AB caried out the experiments. CD performed the data analysis. EF conceived of and designed the study, and drafted the manuscript All auhtors read and approved the manuscript'.}

\competing{The author declares that he has no competing interests.}

\funding{This work was supported by the U.S. Department of Energy,
Office of Science grants No.
DE-FG88ER40388 and DE-SC0012704, and Office of Science, National Quantum
Information Science Research Centers, Co-design Center for Quantum Advantage (${\rm C^2QA}$) under contract number DE-SC0012704.}

\ack{I am indebted to K. Baker, A. Florio, V. Korepin, E. Levin, K. Tu, and T. Ullrich for collaborations and useful discussions.}

%\disclaimer{Insert disclaimer text here if applicable.}

%%%%%%%%%% Insert bibliography here %%%%%%%%%%%%%%

\end{document}